\documentstyle[preprint,aps,epsfig]{revtex}
\begin{document}

\tighten
\preprint{
\vbox{
\hbox{ADP-01-36/T468}
\hbox{JLAB-THY-01-28}
}}

\title{Quark-Hadron Duality and the Nuclear EMC Effect}

\author{W.~Melnitchouk$^1$, K.~Tsushima$^2$\footnote{
	Present address: Department of Physics and Astronomy,
	University of Georgia, Athens, GA 30602, USA}
	and A.~W.~Thomas$^2$}
\address{$^1$ Jefferson Lab, 12000 Jefferson Avenue,
	Newport News, VA 23606, USA}
\address{$^2$ Special Research Centre for the Subatomic Structure
	of Matter, and \\
	Department of Physics and Mathematical Physics,
	Adelaide University, 5005, Australia}

\maketitle

\begin{abstract}
Recent data on polarized proton knockout reactions off $^4$He nuclei 
suggest a small but nonzero modification of proton electromagnetic form
factors in medium.
Using model independent relations derived on the basis of quark-hadron
duality, we relate the medium modification of the form factors to the
modification at large $x$ of the deep-inelastic structure function of a
bound proton.
This places strong constraints on models of the nuclear EMC effect which
assume a large deformation of the intrinsic structure of the nucleon in
medium.
\end{abstract}

\newpage

\section{Introduction}

The modification of hadron properties in the nuclear environment is of
fundamental importance in understanding the implications of QCD for
nuclear physics.
Over the past few years there has been considerable interest in possible
changes to masses, charge radii and other hadron properties in the
nuclear medium.
There is a significant constraint on the possible change in the
``radius'' of a bound nucleon based on $y$-scaling of a bound nucleon
--- especially in $^3$He \cite{yscaling}.
On the other hand, the axial charge of the nucleon is known to be 
suppressed in nuclear $\beta$ decay, and a change in the charge radius
of a bound proton provides a natural suppression of the Coulomb sum rule
\cite{Coulomb}.
One of the most famous nuclear medium effects --- the nuclear EMC effect
\cite{EMC}, or the change in the inclusive deep-inelastic structure
function of a nucleus relative to that of a free nucleon --- has 
stimulated theoretical and experimental efforts for almost two decades
now which seek to understand the dynamics responsible for the change in
the quark-gluon structure of the nucleon in medium \cite{EMCTH}.

The EMC effect illustrates an inherent difficulty in identifying genuine
nuclear quark-gluon effects in a background of purely hadronic physics,
such as conventional nuclear binding and Fermi motion, associated with
the nuclear bound state.
Most features of the nuclear to nucleon structure function ratio can be
(at least qualitatively) understood in terms of conventional nuclear
physics \cite{EMC}.
On the other hand, some of these features can also be attributed to a
modification of the intrinsic nucleon structure function in medium.

Recently the search for evidence of modification of nucleon properties in
medium has been extended to electromagnetic form factors, in polarized
$(\vec e, e' \vec p)$ scattering experiments on $^{16}$O \cite{O16} and
$^4$He \cite{HE4} nuclei.
These experiments measured the ratio of transverse to longitudinal
polarization of the ejected protons, which for a free nucleon is
proportional to the ratio of electric to magnetic elastic form factors
\cite{POLTRANS},
\begin{eqnarray}
{ G_E \over G_M }
&=& - {P'_x \over P'_z} {E + E' \over 2M} \tan(\theta/2)\ ,
\end{eqnarray}
where $P'_x$ and $P'_z$ are the transverse and longitudinal polarization
transfer observables, $E$ and $E'$ the incident and recoil electron
energies, $\theta$ the electron scattering angle, and $M$ the nucleon
mass.
Compared with the more traditional cross section measurements,
polarization transfer experiments provide more sensitive tests of
dynamics, especially of any in-medium changes in the form factor ratios.
The feasibility of this technique was first demonstrated in the
commissioning experiment at Jefferson Lab on $^{16}$O \cite{O16} nuclei
at $Q^2=0.8$~GeV$^2$.
Unfortunately, the errors in this exploratory study were too large to
draw firm conclusions about possible medium modification effects.
In the subsequent experiment at MAMI on $^4$He \cite{HE4} at
$Q^2 \approx 0.4$~GeV$^2$, which had much higher statistics, the
polarization ratio in $^4$He was found to differ by $\approx 10\%$ from
that in hydrogen.

Conventional models using free nucleon form factors and the best
phenomenologically determined optical potentials and bound state wave
functions, as well as relativistic corrections, meson exchange currents,
isobar contributions and final state interactions
\cite{LAGET,KELLY,UDIAS,FOREST}, fail to account for the observed effect
in $^4$He \cite{HE4}.
Indeed, full agreement with the data was only obtained when, in addition
to these standard nuclear corrections, a small change in the structure of
the bound nucleon was taken into account \cite{QMC2,GUICHON,QMC1}.
Regardless of the microscopic origin of the nucleon structure
modification, if there are density dependent effects which modify the
quark substructure of the nucleon, then these should leave traces in a
variety of processes and observables, including structure functions and
form factors.

Of course, one must caution that the study of off-shell nucleon effects is
hampered with difficulties in unambiguously identifying effects associated
with nucleon structure deformation \cite{MST}.
In principle, one can reshuffle strength from off-shell corrections to 
meson exchange currents or interaction terms \cite{OFFSHELL}, so that
``off-shell effects'' can only be identified after specifying a
particular form of the interaction of a nucleon with the surrounding
nuclear medium.
Nevertheless, within a given model of the nucleus, one can study the
capacity to {\em simultaneously} describe form factors and structure
functions as well as static nuclear properties.
It is in this context that we proceed with the discussion of the possible
modifications of nucleon properties in the nuclear medium.

There has recently been considerable interest in the interplay between
form factors and structure functions in the context of quark-hadron
duality.
As observed originally by Bloom and Gilman \cite{BG}, the $F_2$ structure
function measured in inclusive lepton scattering at low $W$ (where $W$ is
the mass of the hadronic final state) generally follows a global scaling
curve which describes high $W$ data, to which the resonance structure
function averages.
Furthermore, the equivalence of the averaged resonance and scaling
structure functions appears to hold for each resonance region, over
restricted intervals of $W$, so that the resonance--scaling duality
also exists locally.
These findings were dramatically confirmed in recent high-precision
measurements of the proton and deuteron $F_2$ structure function at
Jefferson Lab \cite{JLABF2,JLABPAR}, which demonstrated that local
duality works remarkably well for each of the low-lying resonances,
including surprisingly the elastic, to rather low values of $Q^2$.

In this paper we use the concept of quark-hadron duality to relate the
medium dependence of nucleon electromagnetic form factors to the medium
dependence of nucleon structure functions.
To the extent that local duality is a good approximation, these relations
are model independent, and can in fact be used to test the
self-consistency of the models.
We find that the recent form factor data for a proton bound in $^4$He
\cite{HE4} place strong constraints on the medium modification of
inclusive structure functions at large Bjorken-$x$.
In particular, they appear to disfavor models in which the bulk of the
nuclear EMC effect is attributed to deformation of the intrinsic nucleon
structure off-shell -- see e.g. Ref.~\cite{FS}.

In Section~II we discuss the modification of nucleon electromagnetic form
factors as inferred from the recent polarization transfer experiments.
As found in the analysis of the data in Ref.\cite{HE4}, amongst those
models for which predictions were available, the modifications could only
be understood within the context of the quark-meson coupling model
\cite{QMC2,GUICHON,QMC1}.
We therefore use this model to calculate the density dependence of the
bound nucleon electromagnetic form factors.
In Section~III quark-hadron duality is used to relate the observed form
factor modification to that which would be expected in the deep-inelastic
structure functions.
We briefly review the relevant features of Bloom-Gilman duality and
compare the results of models with and without large medium modifications
of the intrinsic nucleon structure.
Finally, in Section~IV we make concluding remarks and discuss
implications of our results for future experiments.

\section{Nuclear Medium Modification of Form Factors}

Let us briefly review the medium modification of the electromagnetic
form factors of the nucleon, as suggested in the recent quasi-elastic
scattering experiment on $^4$He \cite{HE4}.
The data were analyzed using a variety of models, nonrelativistic and
relativistic, based on conventional nucleon-nucleon potentials and
well-established bound state wave functions, including corrections from
meson exchange currents, final state interaction and other effects
\cite{LAGET,KELLY,UDIAS,FOREST}.
The conventional models with the free nucleon form factors could produce
a deviation of at most one half of a percent in the nuclear transverse to
longitudinal ratio, $P'_x/P'_z$, compared with that in hydrogen, although
spinor distortions in fully relativistic calculations were found to
produce an effect of order 2--5\% \cite{UDIAS}.
The observed deviation, which was of order 10\%, could only be explained
by supplementing the conventional nuclear description with the effects
associated with the modification of the nucleon internal structure.
Even though the effect is currently only at the level of 1--2 
standard deviations, it is of considerable interest and importance as 
the first relatively model independent indication of a change in the
internal structure of the nucleon in a nuclear environment.

In the quark-meson coupling (QMC) model \cite{GUICHON,QMC1} the medium
effects arise through the self-consistent coupling of phenomenological
scalar $(\sigma)$ and vector $(\omega_\mu,\rho_\mu)$ meson fields to
confined valence quarks, rather than to the nucleons, as in Quantum
Hadrodynamics \cite{QHD}.
As a result, the internal structure of the bound nucleon is modified by
the surrounding nuclear medium.
The modification of the electromagnetic form factors of the bound nucleon
has been studied using an improved cloudy bag model (CBM) \cite{CBM,DING},
together with the QMC model \cite{QMC2}.
The improved CBM includes a Peierls-Thouless projection to account for
center of mass and recoil corrections, and a Lorentz contraction of the
internal quark wave function.
In this study we calculate the change of the nucleon electromagnetic form
factor in a nuclear medium as in Ref.~\cite{QMC2}.

Because the average nuclear densities for all existing stable nuclei 
heavier than deuterium lie in the range
${1\over 2} \rho_0 \alt \rho \alt \rho_0$, where
$\rho_0 = 0.15$~fm$^{-3}$ is the normal nuclear matter density,
we consider two specific nuclear densities ($\rho = {1\over 2} \rho_0$
and $\rho = \rho_0$) to give the upper and lower bounds for the change
of the electromagnetic form factors (and structure functions at large $x$)
of the bound nucleon.
Furthermore, for the isoscalar $^4$He and $^{16}$O nuclei we neglect the
tiny amount of charge symmetry breaking (due to the Coulomb force and the
$u$ and $d$ current quark mass differences).

The Lagrangian density of the QMC model for symmetric nuclear matter is
given by \cite{GUICHON,QMC1}:
\begin{eqnarray}
{\cal L}_{\rm QMC}
&=& \sum_q
\overline \psi_q (i\gamma^\mu \partial_\mu - m_q) \psi_q\ \theta_V\
 -\ B\ \theta_V\				\nonumber\\
&+& g_\sigma^q\ \overline \psi_q \sigma \psi_q\
 -\ g_\omega^q\ \overline \psi_q \gamma^\mu \omega_\mu \psi_q\
 -\ {1\over 2} m_{\sigma}^2 \sigma^2\
 +\ {1\over 2} m_{\omega}^2 \omega^\mu \omega_\mu\ , 
\end{eqnarray}
where $\psi_q$ is the quark field for a quark flavor $q$, $B$ is the bag
constant, $g_\sigma^q$ and $g_\omega^q$ denote the quark-meson coupling
constants, and $\theta_V$ is a step function equal to unity inside the
confining volume and vanishing outside (the $\rho_\mu$ meson mean field
vanishes in symmetric nuclear matter).
In the mean field approximation, the meson fields are treated as classical
fields, and the quark field inside the bag satisfies the equation of
motion \cite{QMC2}:
\begin{equation}
\left[ i\gamma^\mu\partial_\mu - m_q^*
	- g_\omega^q\ \overline{\omega}\ \gamma^0
\right] \psi_q(x) = 0\ ,
\end{equation}
where $\overline{\sigma}$ and $\overline{\omega}$ denote the constant
mean values of the scalar and the time component of the vector field,
respectively, in symmetric nuclear matter, and
$m_q^* \equiv m_q - g_\sigma^q\ \overline{\sigma}$ is the current quark
mass in the nuclear medium (hereafter we denote the in-medium quantities
by an asterisk $^*$).
The electromagnetic current is given by the sum of the contributions
from the quark core and the pion cloud,
\begin{eqnarray}
\label{current}
j^\mu(x) = \sum_q Q_q e \overline{\psi}_q(x) \gamma^\mu \psi_q(x)
	 - i e [ \pi^\dagger(x) \partial^\mu \pi(x)
		-\pi(x) \partial^\mu \pi^\dagger(x) ]\ ,
\end{eqnarray}
where $Q_q$ is the charge operator for a quark flavor $q$, and $\pi(x)$
destroys a negatively charged (or creates a positively charged) pion.

In the Breit frame the quark core contribution to the electromagnetic
form factors of the bound nucleon is given by~\cite{QMC2}:
\begin{mathletters}
\label{GEMQMC}
\begin{eqnarray}
G_E(Q^2)
&=& \eta^2\ G^{\rm sph}_E(\eta^2 Q^2)\ ,	\\
G_M(Q^2)
&=& \eta^2\ G^{\rm sph}_M(\eta^2 Q^2)\ ,
\end{eqnarray}
\end{mathletters}%
where $Q^2 \equiv -q^2 = \vec q^{\,2}$, and the scaling factor
$\eta = M^*/E^*$, with $E^*=\sqrt{{M^*}^2 + Q^2/4}$ the energy
and $M^*$ the mass of the nucleon in medium,
and $G_{E,M}^{\rm sph}(Q^2)$ are the form factors calculated with the
static spherical bag wave function:
\begin{mathletters}
\begin{eqnarray}
G_E^{\rm sph}(Q^2)
&=& { 1 \over D }
    \int\! d^3r\ j_0(Qr)\ f_q(r)\ K(r)\ ,		\\
G_M^{\rm sph}(Q^2)
&=& { 1 \over D }
    { 2 M \over Q }
    \int\! d^3r\ j_1(Qr)\ \beta_q\
    j_0(\omega_0 r/R)\ j_1(\omega_0 r/R)\ K(r)\ .
\end{eqnarray}
\end{mathletters}%
Here
$f_q(r) = j_0^2(\omega_0 r/R) + \beta_q^2\ j_1^2(\omega_0 r/R)$,
where $R$ is the bag radius, $\omega_0$ the lowest eigenfrequency,
and $\beta_q^2 = (\Omega_q - m_q R)/(\Omega_q + m_q R)$, with
$\Omega_q = \sqrt{ \omega_0^2 + m_q^2 R^2 }$.
The recoil function
$K(r) = \int\! d^3x \, f_q(\vec x) f_q(-\vec x - \vec r)$
accounts for the correlation of the two spectator quarks, and
$D = \int\! d^3r\ f_q(r)\ K(r)$ is the normalization factor.
The scaling factor $\eta$ in the argument of $G_{E,M}^{\rm sph}$ arises
from the coordinate transformation of the struck quark, and the prefactor
in Eqs.(\ref{GEMQMC}) comes from the reduction of the integral measure of
the two spectator quarks in the Breit frame.

The contribution from the pionic cloud is calculated along the lines of
Ref.~\cite{QMC2}.
Although the pion mass would be slightly smaller in the medium than in
free space, the pion field has little effect on the proton form factors,
so that we use $m_\pi^*= m_\pi$.
Furthermore, since the $\Delta$ isobar is treated on the same footing as
the nucleon in the CBM, and because it contains three ground state light
quarks, its mass should vary in a similar manner to that of the nucleon
in the QMC model.
As a first approximation we therefore take the in-medium and free space
$N$--$\Delta$ mass splittings to be approximately equal,
$M_\Delta^* - M^* \simeq M_\Delta - M$.

Including both the quark core and pion cloud contributions, the electric
and magnetic form factors of the free and bound nucleons were calculated
in Ref.\cite{QMC2}.
One finds that the modification of the bound nucleon form factors is
1--2\% for the magnetic and of order 8\% for the electric form factor,
respectively, at normal nuclear matter density ($\rho = \rho_0$), and for
$Q^2 \simeq 0.3$~GeV$^2$, when all form factors are normalized to unity at
$Q^2=0$.
Of course, in the present analysis the absolute value of the proton
magnetic form factor at $Q^2=0$ (the magnetic moment), which is enhanced
in medium, also plays an important role -- as it did in the analysis of
polarized $(\vec e, e'\vec p)$ scattering experiments.
The values of the current quark masses, $m_q \equiv m_u = m_d = 5$~MeV,
and the nucleon bag radius in free space, $R = 0.8$~fm, are the same as
those used in the earlier calculations which reproduce nuclear saturation
properties, and which produced the good agreement with the form factor
data in Ref.\cite{HE4}.
None of the results for nuclear properties, however, depend strongly on
the choice of parameters once the quark-meson coupling constants are
fixed to reproduce the nuclear saturation properties.
As shown in Ref.\cite{QMC1}, the dependence of the properties of finite
nuclei on $m_q$ and $R$ is relatively weak.

The change in the ratio of the electric to magnetic form factors of the
proton,
\begin{eqnarray}
R^p_{EM}(Q^2) &=& { G_E^p(Q^2) \over G_M^p(Q^2) }\ ,
\end{eqnarray}
from free to bound, is illustrated in Fig.~1 for $Q^2$ up to 4~GeV$^2$,
for $\rho = \rho_0$ and $\rho = {1 \over 2} \rho_0$.
Because of charge conservation, the value of $G_E^p$ at $Q^2=0$ remains
unity for any $\rho$.
On the other hand, the proton magnetic moment is enhanced in the nuclear
medium, increasing with $\rho$, so that $R^{p\, *}_{EM} < R^p_{EM}$ at
$Q^2=0$.
In fact, the electric to magnetic ratio is $\sim 5\%$ smaller in medium
than in free space for $\rho={1\over 2}\rho_0$, and $\sim 10\%$ 
smaller for $\rho=\rho_0$.
The effect increases with $Q^2$ out to $\sim 2$~GeV$^2$, where the
bound/free ratio deviates by $\sim 20\%$ from unity.

On the other hand, because nuclear density is not uniform throughout the
nucleus, the $\approx$ 20\% change in the form factors produces only
a few \% effect in the polarization ratio \cite{HE4}.
The experiment not only probes the central region where $\rho$ is
maximal, but also outer regions where $\rho$ is much smaller, so that
integration over the entire nucleus dilutes the effect.
Nevertheless, a form factor modification of this order of magnitude is
needed to explain the observed effect \cite{HE4}.
In the next section we examine the implications of the modification of
the form factors for the medium modification of structure functions at
large $x$.

\section{Quark-Hadron Duality and Nucleon Structure Functions in Medium}

The relationship between form factors and structure functions, or more
generally between inclusive and exclusive processes, has been studied
in a number of contexts over the years.
Drell \& Yan \cite{DY} and West \cite{WEST} pointed out long ago that, 
simply on the basis of scaling arguments, the asymptotic behavior of
elastic electromagnetic form factors as $Q^2 \to \infty$ can be related
to the $x \to 1$ behavior of deep-inelastic structure functions.
In perturbative QCD language, this can be understood in terms of hard
gluon exchange \cite{CM}: deep-inelastic scattering at $x \sim 1$ probes
a highly asymmetric configuration in the nucleon in which one of the
quarks goes far off-shell after the exchange of at least two hard gluons
in the initial state; elastic scattering, on the other hand, requires at
least two gluons in the final state to redistribute the large $Q^2$
absorbed by the recoiling quark \cite{LB}.

More generally, the relationship between resonance (transition) form
factors and the deep-inelastic continuum has been studied in the framework
of quark-hadron, or Bloom-Gilman, duality: the equivalence of the
averaged structure function in the resonance region and the scaling
function which describes high $W$ data.
The recent high precision Jefferson Lab data \cite{JLABF2} on the $F_2$
structure function suggests that the resonance--scaling duality also
exists locally, for each of the low-lying resonances, including
surprisingly the elastic \cite{JLABPAR}, to rather low values of $Q^2$.

In the context of QCD, Bloom-Gilman duality can be understood within
an operator product expansion of moments of structure functions
\cite{RUJ,JI}: the weak $Q^2$ dependence of the low $F_2$ moments can
be interpreted as indicating that higher twist ($1/Q^2$ suppressed)
contributions are either small or cancel.
However, while allowing the duality violations to be identified and
classified according to operators of a certain twist, it does not explain
why some higher twist matrix elements are intrinsically small.

A number of recent studies have attempted to identify the dynamical origin
of Bloom-Gilman duality using simple models of QCD \cite{DOM,IJMV,MODELS}.
It was shown, for instance, that in a harmonic oscillator basis one can
explicitly construct a smooth, scaling structure function from a set of
infinitely narrow resonances \cite{DOM,IJMV}.
Although individual resonance contributions are suppressed by powers of
$1/Q^2$, the number of states accessible increases with $Q^2$ so as to
compensate the fall off, and as $Q^2 \to \infty$ quark-hadron duality
arises from the summation over a complete set of hadronic states.
At lower $Q^2$, however, the appearance of duality could in some cases be
accidental, for example, because of a fortuitous cancellation of
off-diagonal terms in the valence quark charges in the proton
\cite{GOTT,CI,BROD}, allowing a coherent process (exclusive form factors)
to be expressed in terms of incoherent scattering (structure functions).
Whatever the ultimate microscopic origin of Bloom-Gilman duality, for our
purposes it will be sufficient to note the {\em empirical fact} that local
duality is realized in lepton-proton scattering down to
$Q^2 \sim 0.5$~GeV$^2$ at the 10-20\% level for the lowest moments of the
structure function.
In other words, here we are not concerned about {\em why} duality works,
but rather {\em that} it works.

Motivated by the experimental verification of local duality, one can use
measured structure functions in the resonance region to directly extract
elastic form factors \cite{RUJ}.
Conversely, empirical electromagnetic form factors at large $Q^2$ can 
be used to predict the $x \to 1$ behavior of deep-inelastic structure
functions \cite{BG,CM,ELDUAL,QNP}.
The assumption of local duality for the elastic case implies that the area
under the elastic peak at a given $Q^2$ is equivalent to the area under
the scaling function, at much larger $Q^2$, when integrated from the pion
threshold to the elastic point \cite{BG}.
Using the local duality hypothesis, de R\'ujula et al. \cite{RUJ}, and
more recently Ent et al. \cite{JLABPAR}, extracted the proton's magnetic
form factor from resonance data on the $F_2$ structure function at large
$x$, finding agreement to better than 30\% over a large range of $Q^2$
($0.5 \alt Q^2 \alt 5$~GeV$^2$).
In the region $Q^2 \sim 1$--2~GeV$^2$ the agreement was at the $\sim 10\%$
level.
An alternative parameterization of $F_2$ was suggested in
Ref.\cite{SIMULA}, which because of a different behavior in the unmeasured
region $\xi \agt 0.86$, where $\xi = 2 x / (1 + \sqrt{1 + x^2/\tau})$ is
the Nachtmann variable, with $\tau = Q^2/4M^2$, led to larger differences
at $Q^2 \agt 4$~GeV$^2$.
However, at $Q^2 \sim 1$~GeV$^2$ the agreement with the form factor data
was even better here.
As pointed out in Ref.\cite{REPLY}, data at larger $\xi$ are needed to
constrain the structure function parameterization, and reliably extract
the form factor at larger $Q^2$.
Furthermore, since we will be interested in {\em ratios} of form factors
and structure functions only, what is more relevant for our analysis is
not the degree to which local duality holds for the {\em absolute}
structure functions, but rather the {\em relative} change in the duality
approximation between free and bound protons.

Applying the argument in reverse, one can formally differentiate the local
elastic duality relation \cite{BG} with respect to $Q^2$ to express the
scaling functions, evaluated at threshold, 
$x = x_{\rm th} = Q^2 / (W^2_{\rm th} - M^2 + Q^2)$, with
$W_{\rm th} = M + m_\pi$, in terms of $Q^2$ derivatives of elastic form
factors.
In Refs.\cite{BG,ELDUAL} the $x \to 1$ behavior of the neutron to proton
structure function ratio was extracted from data on the elastic
electromagnetic form factors.
(Nucleon structure functions in the $x \sim 1$ region are important as
they reflect mechanisms for the breaking of spin-flavor SU(6) symmetry in
the nucleon \cite{MT}.)
Extending this to the case of bound nucleons, one finds that as
$Q^2 \to \infty$ the ratio of bound to free proton structure functions is:
\begin{eqnarray}
\label{SFdual}
{ F_2^{p\, *} \over F_2^p }
&\to& { dG_M^{p\, *\, 2} / dQ^2 \over
        dG_M^{p\, 2 } / dQ^2 }\ .
\end{eqnarray}
At finite $Q^2$ there are corrections to Eq.(\ref{SFdual}) arising from
$G_E^p$ and its derivatives, as discussed in Ref.\cite{ELDUAL}.
(In this analysis we use the full, $Q^2$ dependent expressions
\cite{ELDUAL,QNP}.)
Note that in the nuclear medium, the value of $x$ at which the pion
threshold arises is shifted:
\begin{eqnarray}
x_{\rm th} &\to& x^*_{\rm th}\
=\ \left( { m_\pi ( 2 M + m_\pi ) + Q^2 \over
	    m_\pi (2 (M^* + V) + m_\pi ) + Q^2 }
   \right) x_{\rm th}\ ,
\end{eqnarray}
where $V = 3 g^q_\omega\ \bar{\omega}$ is the vector potential felt by
the nucleon and (consistent with chiral expectations and phenomenological
constraints) we have set $m_\pi^* = m_\pi$.
However, the difference between $x_{\rm th}$ and $x^*_{\rm th}$ has a
negligible effect on the results for most values of $x$ considered.

Using the duality relations between electromagnetic form factors and
structure functions, in Fig.~2 we plot the ratio $F_2^{p *}/F_2^p$ as
a function of $x$, with $x$ evaluated at threshold, $x = x_{\rm th}$
(solid lines).
Note that at threshold the range of $Q^2$ spanned between $x=0.5$ and
$x=0.8$ is $Q^2 \approx 0.3$--1.1~GeV$^2$.
Over the range $0.5 \alt x \alt 0.75$ the effect is almost negligible,
with the deviation of the ratio from unity being $\alt 1\%$ for
$\rho={1\over 2}\rho_0$ and $\alt 2\%$ for $\rho=\rho_0$.
For $x \agt 0.8$ the effect increases to $\sim 5\%$, although, since
larger $x$ corresponds to larger $Q^2$, the analysis in terms of the QMC
model is less reliable here.
However, in the region where the analysis can be considered reliable, the
results based on the bound nucleon form factors inferred from the  
polarization transfer data \cite{HE4} and local duality imply that the
nucleon structure function undergoes very little modification in medium.

It is instructive to contrast this result with models of the EMC effect
in which there is a large medium modification of nucleon structure.
For example, let us consider the model of Ref.\cite{FS}, where it is
assumed that for large $x$ the dominant contribution to the structure
function is given by the point-like configurations (PLC) of partons which
interact weakly with the other nucleons.
The suppression of this component in a bound nucleon is assumed to be the
main source of the EMC effect.
This model represents one of the extreme possibilities that the EMC
effect is solely the result of deformation of the wave function of bound
nucleons, without attributing any contribution to nuclear pions or other
effects associated with nuclear binding \cite{MSS}.
Given that this model has been so successfully applied to describe the
nuclear EMC effect, it is clearly important to examine its consequences
elsewhere.

The deformation of the bound nucleon structure function in the PLC
suppression model is governed by the function \cite{FS}:
\begin{eqnarray}
\label{delta}
\delta(k) &=& 1 - 2 (k^2/2M + \epsilon_A)/\Delta E_A\ ,
\end{eqnarray}
where $k$ is the bound nucleon momentum, $\epsilon_A$ is the nuclear
binding energy, and $\Delta E_A \sim 0.3$--0.6~GeV is a nucleon
excitation energy in the nucleus.
For $x \agt 0.6$ the ratio of bound to free nucleon structure functions
is then given by \cite{FS}:
\begin{eqnarray}
\label{plc}
{ F_2^{N\, *}(k, x) \over F_2^N(x) }
&=& \delta(k)\ .
\end{eqnarray}
The $x$ dependence of the suppression effect is based on the assumption
that the point-like configuration contribution in the nucleon wave
function is negligible at $x \alt 0.3$ ($F_2^{N\, *}/F_2^N = 1$), and for
$0.3 \alt x \alt 0.6$ one linearly interpolates between these values
\cite{FS}.
The results for $^4$He and $^{16}$O are shown in Fig.~2 (dashed lines)
for the average values of nucleon momentum, $\langle k^2 \rangle$, in
each nucleus.
The effect is a suppression of order 20\% in the ratio
$F_2^{N\, *}/F_2^N$ for $x \sim 0.6$--0.7.
In contrast, the ratios extracted on the basis of duality, using the QMC
model constrained by the $^4$He polarization transfer data \cite{HE4},
show almost no suppression ($\alt 1$--2\%) in this region.
Thus, for $^4$He, the effect in the PLC suppression model is an order
of magnitude too large at $x \sim 0.6$, and has the opposite sign for
$x \agt 0.65$.

Although the results extracted from the polarization transfer
measurements \cite{HE4} rely on the assumption of local duality, we
stress that the corrections to duality have been found to be typically
less than 20\% for $0.5 \alt Q^2 \alt 2$~GeV$^2$ \cite{JLABF2,SIMULA}.
The results therefore appear to rule out large bound structure function
modifications, such as those assumed in the point-like configuration
suppression model \cite{FS}, and instead point to a small medium
modification of the intrinsic nucleon structure, which is complemented
by standard many-body nuclear effects.

As a consistency check on the analysis, one can also examine the change
in the form factor of a bound nucleon that would be implied by the
corresponding change in the structure function in medium.
Namely, from the local duality relation \cite{RUJ,QNP}:
\begin{eqnarray}
\label{elint}
\left[ G_M^p(Q^2) \right]^2
&\approx& { 2 - \xi_0 \over \xi_0^2 }
	  { (1 + \tau) \over (1/\mu_p^2 + \tau) }
	  \int_{\xi_{\rm th}}^1 d\xi\ F_2^p(\xi)\ ,
\end{eqnarray}
one can extract the magnetic form factor by integrating the $F_2^p$
structure function over $\xi$
between threshold, $\xi = \xi_{\rm th}$, and $\xi=1$.
Here $\xi_0 = \xi(x=1)$, and $\mu_p$ is the proton magnetic moment.
In Fig.~3 we show the PLC model predictions for the ratio of the magnetic
form factor of a proton bound in $^4$He to that in vacuum, derived from
Eqs.(\ref{plc}) and (\ref{elint}), using the parameterization for
$F_2^p(\xi)$ from Ref.\cite{JLABPAR}, and an estimate for the in-medium
value of $\mu_p^*$ from Ref.\cite{QMC2}.
Taking the average nucleon momentum in the $^4$He nucleus,
$k = \langle k \rangle$, the result is a suppression of about 20\% in
the ratio $G^{p *}_M/G^p_M$ at $Q^2 \sim 1$--2~GeV$^2$ (solid curve).
Since the structure function suppression in the PLC model depends on the
nucleon momentum (Eq.(\ref{delta})), we also show the resulting form
factor ratio for a momentum typical in the $(\vec e, e' \vec p)$
experiment, $k = 50$~MeV (long dashed).
As expected, the effect is reduced, however, it is still of the order
15\% since the suppression also depends on the binding energy, as well
as on the nucleon mass, which changes with density rather than with momentum.
In contrast, the QMC calculation, which is consistent with the MAMI
$^4$He quasi-elastic data, produces a ratio which is typically 5--10\%
{\it larger} than unity (dashed).
Without a very large compensating change in the in-medium electric form
factor of the proton (which seems to be excluded by $y$-scaling
constraints), the behavior of the magnetic form factor implied by the
PLC model $+$ duality would produce a large {\em enhancement} of the
polarization transfer ratio, rather than the observed small suppression
\cite{HE4}.

\section{Conclusion}

In this paper we have examined the consequences of quark-hadron duality
applied to nucleons in the nuclear medium.
Utilizing the experimental results \cite{HE4} from polarized proton
knockout reactions off $^4$He nuclei, which suggest a small but nonzero
modification of the proton electromagnetic form factors in medium, we
use local duality to relate {\em model-independently} the medium modified
form factors to the change in the intrinsic structure function of a bound
proton.

The analysis in Ref.\cite{HE4} found that, compared with conventional
nuclear calculations, the medium modifications observed in the $^4$He
data could only be described within models which allowed a small
modification of the nucleon form factors in medium, such as the
quark-meson coupling model \cite{QMC2,GUICHON,QMC1} (see also
Ref.\cite{CELENZA}).
In the context of the QMC model, the change in nucleon form factors
allowed by the data imply a modification of the in-medium structure
function of $\alt 1$--2\% at $0.5 \alt x \alt 0.75$ for all nuclear
densities between nuclear matter density, $\rho=\rho_0$, and
$\rho={1\over 2}\rho_0$.
While the results rely on the validity of quark-hadron duality, the
empirical evidence suggests that for low moments of the proton's $F_2$
structure function the duality violations due to higher twist corrections
are $\alt 20\%$ for $Q^2 \agt 0.5$~GeV$^2$ \cite{JLABF2}, and decrease
with increasing $Q^2$.

The results place rather strong constraints on models of the nuclear EMC
effect, especially on models which assume that the EMC effect arises from
a large deformation of the nucleon structure in medium.
For example, we find that the PLC suppression model \cite{FS} predicts
an effect which is about an order of magnitude larger than that allowed
by the data \cite{HE4}, and has a different sign.
The findings therefore appear to disfavor models with large medium
modifications of structure functions as viable explanations for the
nuclear EMC effect, although it would be desirable to have more data
on a variety of nuclei and in different kinematical regions.
The recently completed Jefferson Lab $^4$He polarization transfer
experiment, which covered a large range of $Q^2$, between 0.5~GeV$^2$
and 2.6~GeV$^2$ \cite{E93049}, should provide valuable additional
information.
Preliminary results \cite{HE4PRELIM} indicate that the lowest $Q^2$ point
is in very good agreement with the Mainz $Q^2=0.4$~GeV$^2$ data point,
which provides further support for the QMC description.
In addition, the proposed Jefferson Lab experiment on $^{16}$O
\cite{P01013} at $Q^2=0.8$~GeV$^2$, which would make use of other, high
precision cross section data at this momentum transfer, would have
about 15 times the statistics of the original commissioning experiment
\cite{O16}.
This would enable a more thorough comparison of the medium dependence
of form factors and structure functions for different nuclei.

These results have other important practical ramifications.
For instance, the PLC suppression model was used recently \cite{SSS} to
argue that the EMC effects in $^3$He and $^3$H differ significantly at
large $x$, in contrast to calculations \cite{AFNAN,PACE} based on
conventional nuclear physics using well-established bound state wave
functions which show only small differences.
Based on the findings presented here, one would conclude that the
conventional nuclear physics description of the $^3$He/$^3$H system
should indeed be a reliable starting point for nuclear structure function
calculations, as the available evidence suggests little room for large
off-shell corrections.
Finally, let us stress that quark-hadron duality is a powerful tool with
which to simultaneously study the medium dependence of both exclusive
and inclusive observables, and thus provides an extremely valuable guide
towards a consistent picture of the effects of the nuclear environment on
nucleon substructure.

\acknowledgements

We would like to thank S.~Strauch and S.~Dieterich for helpful
discussions and communications, D.H.~Lu for providing the computer codes
for the improved bag model form factor calculations, and J.~Arrington
for a helpful discussion.
This work was supported by the Australian Research Council, and the U.S.
Department of Energy contract \mbox{DE-AC05-84ER40150}, under which the
Southeastern Universities Research Association (SURA) operates the
Thomas Jefferson National Accelerator Facility (Jefferson Lab). \\


\begin{figure}
\epsfig{figure=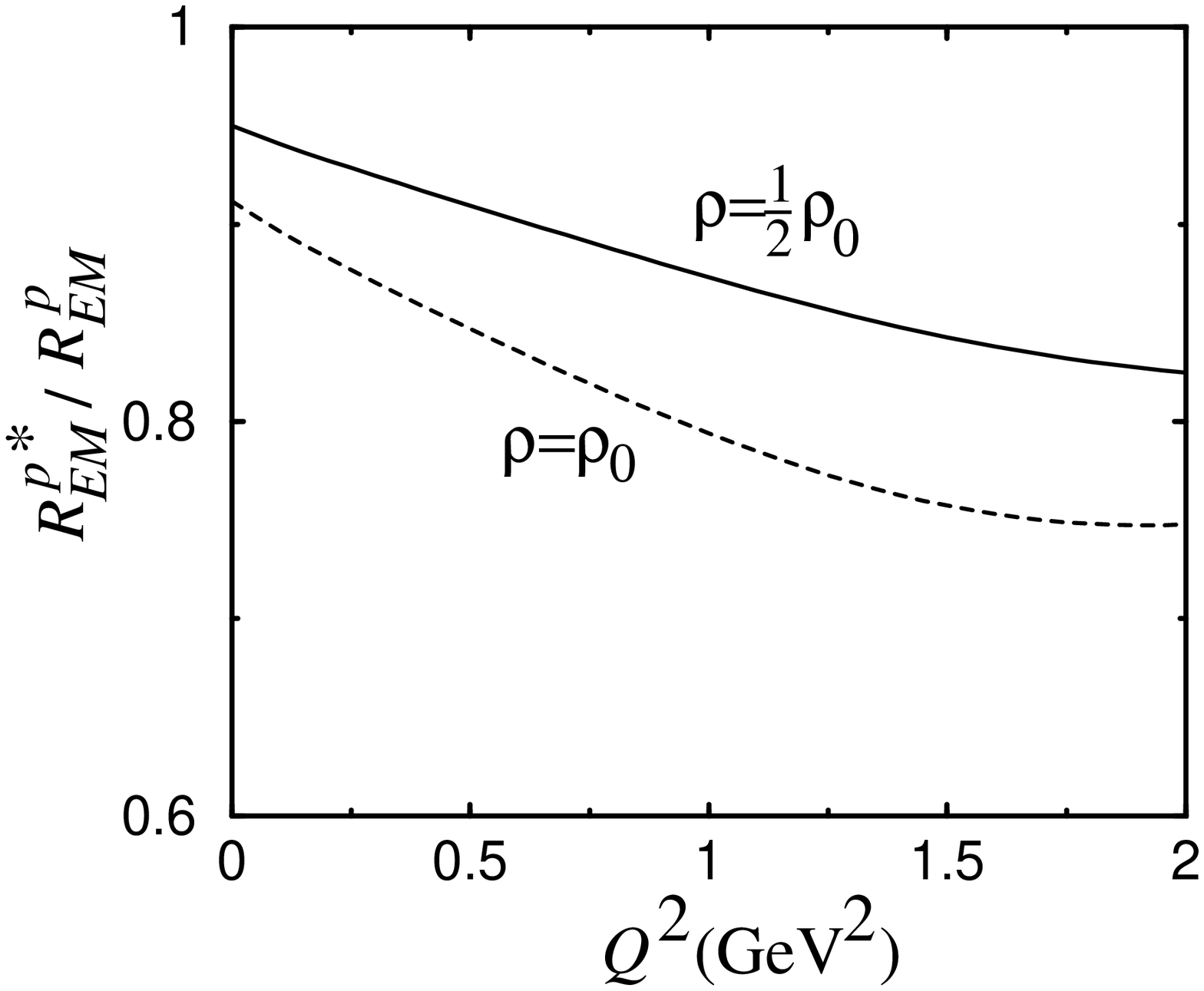,height=11cm}\\
\caption{Comparison of the ratio of electric to magnetic form factors
	of the proton, $R^p_{EM} = G^p_E/G^p_M$, in medium to that
	in free space in the QMC model \protect\cite{QMC2}.
	The bound proton form factors are calculated at nuclear matter
	density, $\rho=\rho_0$ (dashed), and at $\rho={1 \over 2}\rho_0$
	(solid).}
\end{figure}

\newpage

\begin{figure}
\epsfig{figure=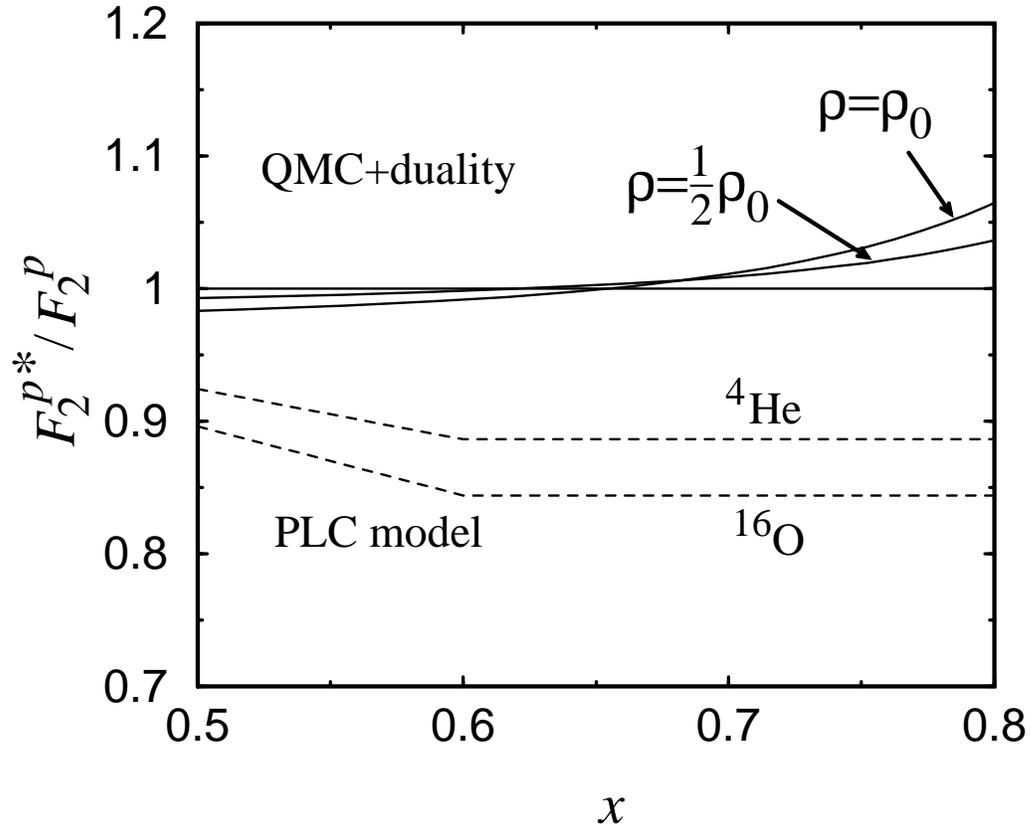,height=11cm}\\
\caption{In-medium to free proton $F_2$ structure function ratio as a
	function of $x$ at threshold, $x=x_{\rm th}$, extracted from the
	polarization transfer data \protect\cite{HE4} within the QMC model
	and local duality, at nuclear matter density, $\rho = \rho_0$,
	and at $\rho = {1 \over 2}\rho_0$ (solid).
	For comparison the results of the PLC suppression model
	\protect\cite{FS} are shown for $^4$He and $^{16}$O
	(dashed).}
\end{figure}

\newpage
        
\begin{figure}
\epsfig{figure=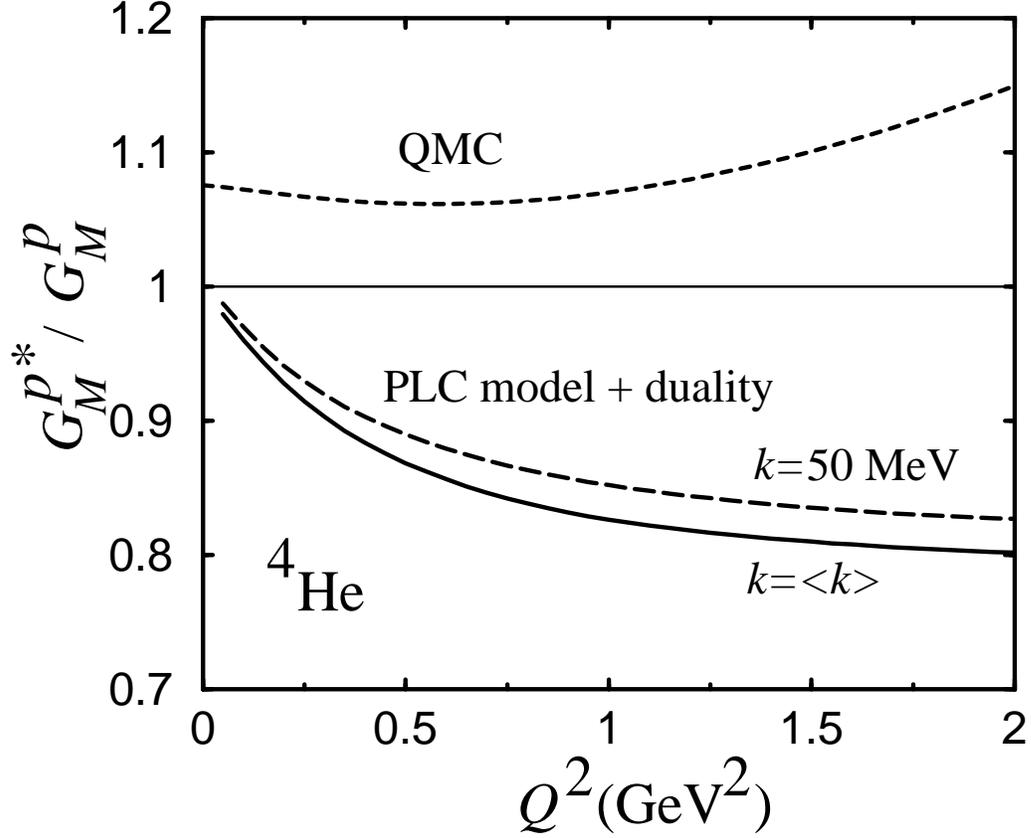,height=11cm}\\
\caption{Ratio of in-medium to free proton magnetic form factors,
	extracted from the PLC suppression model \protect\cite{FS}
	for the EMC ratio in $^4$He, using the $F_2^p$ data from
	Refs.\protect\cite{JLABF2,JLABPAR} and local duality, for
	$k=\langle k \rangle$ (solid) and $k = 50$~MeV (long-dashed).
	The QMC model prediction (short-dashed) is shown for comparison.}
\end{figure}

\end{document}